\documentstyle[twoside,fleqn,espcrc2]{article}
\def\be{\begin{equation}}
\def\ee{\end{equation}}
\def\bea{\begin{eqnarray}}
\def\eea{\end{eqnarray}}

\newcommand{\AmS}{{\protect\the\textfont2
  A\kern-.1667em\lower.5ex\hbox{M}\kern-.125emS}}

\title{Heavy-light physics with NRQCD}

\author{UKQCD collaboration presented by C. T. H. Davies\address{Department of
Physics and Astronomy,
        University of Glasgow, \\
        Glasgow G12 8QQ, UK}%
         }

\begin{document}

\begin{abstract}
First results are obtained for B mesons using a heavy propagator
calculated using NRQCD, and a light Wilson propagator. Results from
13 quenched configurations of size $16^3 \times 48$ at $\beta$ = 6.0
give a value for $f_{B}$ of less than 200 MeV and a $B^{*}-B$ splitting of
32(8) MeV.
Superior signal/noise behaviour is observed over static propagators on the
same configurations. No
extrapolation to the $b$ mass for the heavy quark is required.
\end{abstract}

\maketitle

\section{INTRODUCTION}

B physics is an exciting area of topical interest.
Results for B decay constants and mixing
parameters are urgently needed by the experimental community and
can best be provided by calculations in lattice QCD.
There has been some controversy, however, in interpreting the
lattice results. The static approximation gives large values for
$f_B$ with a large spread from group to group~\cite{bernard}. The poor
signal/noise properties of this approximation presumably give rise
to this variation. Calculations to date using Wilson propagators
for the heavy quark have required an extrapolation in the
heavy quark mass to the $b$.
The use of NRQCD for the heavy quark gives a
result with good signal/noise properties
right at the $b$ mass and the method is
presented here.

NRQCD is an effective field theory, appropriate to heavy quark physics,
which has had great success to date in
the simulation of heavy-heavy systems, both $b\overline{b}$ and
$c\overline{c}$~\cite{sloan,lidsey}.

In that case the NRQCD action is viewed as an expansion in powers of
$v^{2}/c^{2}$ where $v$ is a typical velocity inside
the heavy-heavy system. Simulations have been done recently by the
NRQCD collaboration using leading and next-to-leading terms in
this expansion~\cite{sloan,lidsey}. The
coefficients of the sub-leading terms can be taken at their
tree-level values when the input gauge fields are
transformed as
\begin{equation}
U_{\mu} \rightarrow \frac {U_{\mu}} {u_0} .
\end{equation}
$u_0$ is taken as the fourth root of the average plaquette~\cite{nakhleh}.
What this transformation does is to include in each
term the radiative tadpole corrections to all orders in perturbation
theory.

The success of the spectrum calculated~\cite{sloan,lidsey} gives us confidence
in the action
and in the coefficients of each term. That the associated
perturbation theory is well behaved and understood is shown
by a study of the zero and finite momentum energies of the
$\Upsilon$ as a function of bare quark mass~\cite{sloan}.
 The bare quark mass that should
appear in the action can then be fixed by setting the kinetic mass of a known
state, say the $\Upsilon$, to its experimental value.
This gives a bare quark mass for the $b$, $M_b a$ = 1.7 at
 $\beta$ = 6.0~\cite{sloan}.

To study heavy-light systems we can use the same NRQCD action for the
heavy quark as has been used in the heavy-heavy
 simulations~\cite{thacker}. There
are then no parameters to be fixed at all and $all$ masses are
obtained as predictions.

For heavy-light systems the importance of different terms in the
action changes. The momentum scale inside the mesons is set
now by the dynamics of the light quark and, to leading order, is
independent of $M$, the mass of the heavy quark. The
importance of terms in the NRQCD action then depends only on
their power of $1/M$. Terms at $1/M^2$ which were
included at next-to-leading order in a heavy-heavy simulation
are effectively suppressed.
We can neglect them when working to $1/M$ for a heavy-light
simulation~\cite{thacker}.

The results described here use an NRQCD action for the heavy quark
of the following form
\bea \label{LNRQCD}
{\cal{L}}_{NRQCD} = -\Psi^{\dagger}D_{t}\Psi +
\Psi^{\dagger}\frac{D^{2}}{2M}\Psi + \nonumber \\
\frac{g}{2M}\Psi^{\dagger}\sigma.B\Psi
\eea

The heavy quark propagators are calculated with the
usual evolution equation~\cite{sloan} and with $U_{\mu}$ fields transformed
as in equation 1. The bare heavy quark mass was taken as 2.0 to
simulate a $b$ quark on
lattices at $\beta$ = 6.0. Subsequent work on the
heavy-heavy spectrum as decribed above~\cite{sloan} shows this value for
the mass to
be 15\% high. This initial study used 13 $16^{3}\times 48$
lattices fixed to Coulomb gauge.

The heavy quark propagators were combined with light Wilson
propagators at two $\kappa_l$ values, 0.154 and 0.155. $\kappa_c$ is
determined at this $\beta$ value to be 0.157~\cite{APE}.
The light quark propagators were calculated from a
$\delta$ function source and given a $\sqrt{2\kappa_l}$ rescaling.

The heavy quark propagators were calculated both from a $\delta$
function source and
from an exponentially smeared source ($\rm{exp}$$(-r/r_0)$, with
$r_0$ = 3.0).  There was one source per configuration.
By combining either smeared or local heavy propagators with
local light propagators we are
able to study local-local, smeared-local and smeared-smeared
correlation functions for the B meson.
We also calculated heavy quark propagators in the
static approximation (that is, using equation 2 with all
$1/M$ terms switched off) from the same smeared sources.
This enables us to directly compare the static and NRQCD
signal/noise behaviour.

\section{RESULTS}

\subsection{$f_B$}

To make a $B$ meson we used the na\"{\i}ve lowest order vertex operator,
i.e. we combined the 2-component heavy antiquark propagator with
the 2 upper spins of the light quark propagator. $1/M$ effects in the
vertex operator were
ignored at this stage.
The zero-momentum B meson correlation function then
has an exponential fall-off
given by its energy in lattice units and an amplitude
related to the decay constant, $f_B$.
For the NRQCD heavy quarks the signal/noise is such
that it is possible to fit the
local-local correlation function, see Figure 1.
This has previously been shown not to be possible for
static heavy quarks~\cite{bouchaud}.
Fitting to a single exponential from time slices 12-48
and using an svd algorithm with cut-off 1.0e-03 to
invert the covariance matrix, we obtain at $\kappa_l$ = 0.155:
\begin{eqnarray}
E = 0.48(1), Z_L = 0.11(2), \nonumber \\
\chi^{2}/dof = 1.7/4, Q = 0.8.
\end{eqnarray}
$Z_L$ is the square root of the amplitude. $\sqrt{2}Z_L$ is
equal to $f_B\sqrt{M_B}$ in the $f_{\pi}$ = 132 MeV
convention, modulo a lattice-to-continuum
renormalisation constant. This renormalisation
constant has only been calculated for an NRQCD action
with no $\sigma.B$ term and with no $u_0$ transformation
of the $U_{\mu}$~\cite{thacker2}. A fairly large $\cal{O}$$(g^{2})$
contribution was found in that case, but a major component
of it was the wave function renormalisation of the heavy quark.
This is much reduced when tadpole effects are taken care of
using eq. 1~\cite{nakhleh,morning}. It seems likely that the
renormalistion constant for the action used here is then
much closer to 1.

Extrapolating linearly to $\kappa_c$ for the light quark,
we obtain 0.16(3)/$Z_A$ for $f_B\sqrt{M_B}$ in lattice
units. This agrees with other UKQCD results in this region~\cite{ukqcd},
for reasonable values of $Z_A$.

The predicted value for $f_B$ is then less than 200 MeV, using an
inverse lattice spacing at $\beta=6.0$ of 2.0 GeV. In the
quenched approximation the value for the lattice spacing
depends on the scale at which it is being determined; the
$a^{-1}$ value from $b\overline{b}$ splittings being
considerably larger than that from light hadrons~\cite{sloan}.
The scales appropriate to heavy-light mesons are
presumably much more like those of light
hadrons, so we use an $a^{-1}$ from those calculations, as
above.

\begin{figure}[htb]
\vspace{60mm}
\caption{The effective mass for the local-local correlation function
for the $B$ meson using an NRQCD propagator for the heavy quark.}
\end{figure}

The static calculation requires the calculation of both
smeared-local and smeared-smeared correlation functions.
The smeared-local correlation function is much noisier than the
NRQCD smeared-local function (see Figure 2), but the smeared-smeared
correlation function is only slightly noisier than its
NRQCD counterpart.

The poor noise properties of the local static
correlation functions are because of the absence of a mass
in the heavy-heavy channel that appears in the noise~\cite{lepage}.
Smearing has the effect of producing such a mass by
effectively introducing the heavy quark potential at
non-zero $R$ into the heavy-heavy cross-terms. This is a
wholly desirable effect for the static smeared-local
correlation function and the noise is much reduced. For
NRQCD the non-locality of the source has the effect of
increasing the noise over the local-local function,
although an earlier plateau is seen. In both cases
smearing at the sink introduces more noise than having a local sink.

We can fit a single exponential to the ratio of the square of the
smeared-local to the smeared-smeared correlation functions for the
static case. The amplitude is then the effective local-local
amplitude. We obtain at $\kappa_l$ = 0.155
\begin{eqnarray}
E = 0.47(1), Z_L = 0.21(1), \nonumber \\
\chi^{2}/dof = 6.7/13, Q = 0.9.
\end{eqnarray}
Of necessity there was a more restricted range in time (15 - 25)
than for the NRQCD fit. We believe that this accounts for the
fact that the errors are not significantly worse.

The value for $f\sqrt{M}$ is 0.27(2)/$Z_A$ after
linear extrapolation to the chiral limit in $\kappa_l$. This
is in agreement with other UKQCD static results at
$\beta$ = 6.0~\cite{ukqcd}, obtained using the improved clover
action for the light quarks, although the $Z_A$ values will
be slightly different (by 10\%) in the two cases.
The value for $f_B$ is around 300 MeV, much larger than for
NRQCD heavy qaurks.

\subsection{$M_B$}

Because of the form of the NRQCD action, the energy measured
for a zero-momentum correlation function must be adjusted
to give the mass of the meson. This adjustment by terms which
are perturbatively calculable works very well for heavy-heavy
mesons~\cite{sloan}. For the B meson we have
\begin{equation}
M_B = Z_M M_b - E_0 + E .
\end{equation}
All quantities are in lattice units. $Z_M$ is the
heavy quark mass renormalistion and $E_0$ is the energy
shift. Using $E_0$ = 0.22 and $Z_M$ = 1.14 ~\cite{sloan} gives
$M_B$ = 2.46(5) in lattice units after extrapolation
to the light quark chiral limit. Multiplication by
$a^{-1}$ is the major source of error, giving a result
of 4.9(4) GeV, certainly consistent with experiment (5.3GeV).

Another determination of $M_B$ comes from the
finite momentum propagators. A kinetic mass
can be extracted from the ratio of the energies of
lowest non-zero momentum to the zero-momentum.
With only 13 configurations the results are too noisy
to give a value from this method.

\subsection{$B^{*}-B$}

 From the ratio of the local-local NRQCD correlators for
the B* and the B we can extract a value for
the splitting between B* and B. A single
exponential fit at $\kappa_l$ = 0.155
to the ratio from time slices 5-20
gives
\begin{eqnarray}
E = 0.016(3), A = 1.02(2), \nonumber \\
\chi^{2}/dof = 1.9/3, Q = 0.6.
\end{eqnarray}
Both values of $\kappa_l$ give the same splitting
so no chiral extraploation can be done.
Taking the value above and $a^{-1}$ = 2.0(1) GeV gives
a splitting of 32(8) MeV, rather lower than
the experimental value of 46 MeV.
This may be a reflection of the rather low
quark mass that we have used. If the splitting
is proportional to $1/M_b$, then a quark mass of
1.7 rather than 2 would give a splitting of 38(8) MeV,
consistent with experiment. Values obtained using
heavy Wilson fermions have been much lower than this~\cite{ukqcd2}.

\begin{figure}[htb]
\vspace{60mm}
\caption{A comparison of the effective masses for the smeared-local
correlation function for a $B$ meson using NRQCD propagators (crosses)
and static propagators (circles) for the heavy quark.}
\end{figure}

\subsection{$\Lambda_b$}

We have calculated the NRQCD smeared(h)-local(l)
correlation functions for the $\Lambda_b$. The effective
mass plots are rather noisy, and the mass depends quite
strongly on $\kappa_l$. Chiral extrapolation gives
a splitting between the $\Lambda_b$ and the $B$ of
0.28(8) in lattice units. Converting to
physical units gives the rather imprecise value at
present of
60(20) MeV. The experimental result is 36(4) MeV.

\section{CONCLUSIONS}

The use of NRQCD for heavy quark propagators can
give useful results for heavy-light mesons. Once
the terms in the action are fixed from the very
precise heavy-heavy calculations that are possible,
there are no free parameters left.
Correlation functions for the B meson can be
calculated directly at the $b$ mass (fixed from
$\Upsilon$ spectroscopy) and they have good
signal/noise properties.

This preliminary calculation on a small number
of configurations at $\beta$ = 6.0 finds
a value for $f_B$ consistent with others using
heavy Wilson propagators. This confirms from another
source the existent of relatively large $1/M$
terms in $f\sqrt{M}$ at the the $B$. The usefulness of the
 approach here is
that a direct comparison with static results
calculated using the same smearing functions on the
same configurations is possible.

A value for $M_B$ can be calculated since all
parameters in the action are already fixed. This
agrees with experiment. In addition the value for
the $B^*-B$ splitting looks promising. It is
reasonably close to experiment unlike values obtained
up to now using heavy Wilson fermions.

\begin{flushleft}
{\bf Acknowledgements}

This work was carried out on a Thinking Machines CM-200 supported by
SERC, Scottish Enterprise and the Information Systems Committee
of the UFC. I am grateful to my colleagues in the NRQCD collaboration
for useful discussions.
\end{flushleft}

\end{document}